\documentclass{article}

\usepackage{authblk}
\usepackage[utf8]{inputenc} 
\usepackage[T1]{fontenc}    
\usepackage{hyperref}       
\usepackage{url}            
\usepackage{booktabs}       
\usepackage{amsfonts}       
\usepackage{nicefrac}       
\usepackage{microtype}      
\usepackage{graphicx}
\usepackage{fancyhdr,graphicx,amsmath,amssymb}
\usepackage{fixltx2e}
\usepackage[ruled,vlined]{algorithm2e}
\graphicspath{ {./images/} }

\DeclareMathOperator*{\argmin}{arg\,min}
\newcommand{\norm}[1]{\left\lVert#1\right\rVert}

\date{}

\title{Automatic Differentiation for All Photons Imaging to See Inside Volumetric Scattering Media}

\author[1]{Tomohiro Maeda\thanks{Corresponding Author: tomotomo@mit.edu}}
\author[2]{Ankit Ranjan}
\author[1]{Ramesh Raskar}
\affil[1]{Media Lab, Massachusetts Institute of Technology}
\affil[2]{Nuffield Department of Surgical Sciences, University of Oxford}

\begin{document}
\maketitle
\begin{abstract}
Imaging through dense scattering media — such as biological tissue, fog, and smoke — has applications in the medical and robotics fields. We propose a new framework using automatic differentiation for All Photons Imaging through homogeneous scattering media with unknown optical properties for non-invasive sensing and diagnostics. We overcome the need for the imaging target to be visible to the illumination source in All Photons Imaging, enabling practical and non-invasive imaging through turbid media with a simple optical setup. Our method does not require calibration to acquire the sensor position or optical properties of the media. 
\end{abstract}


\section*{Introduction}
\begin{figure}
        \centering
        \includegraphics[width=0.8\linewidth]{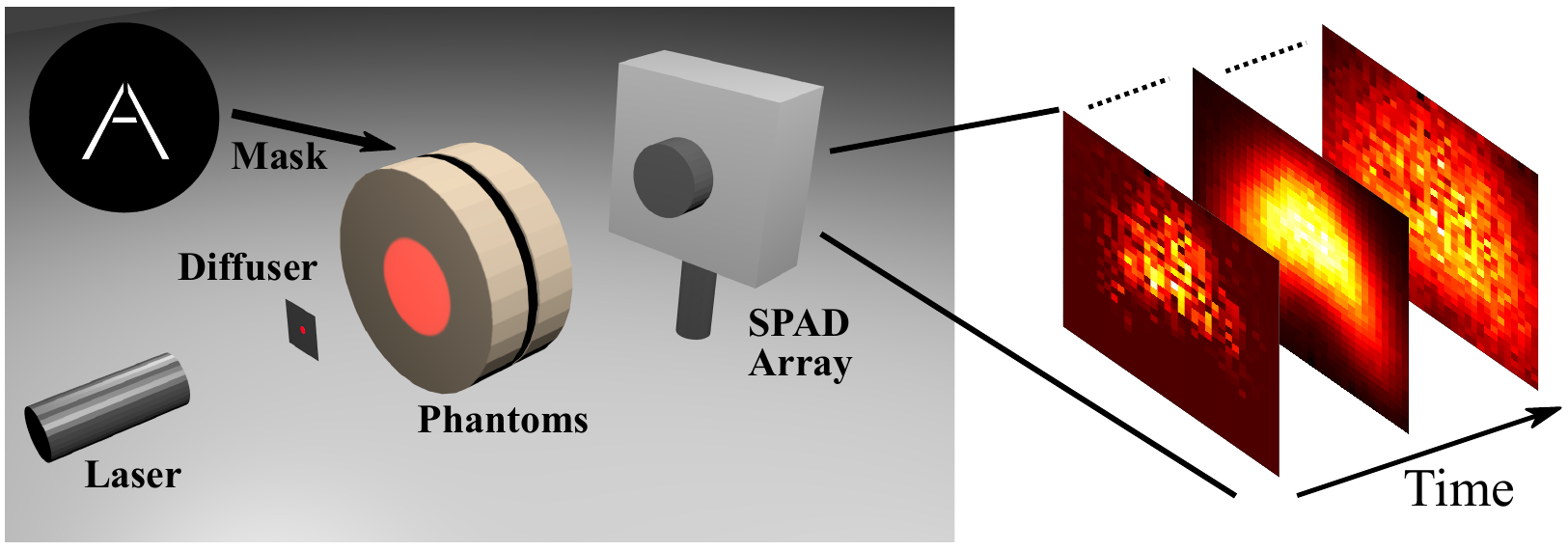}
        \caption{\textbf{Imaging system to see through scattering media.} A single-photon avalanche diode (SPAD) array captures 3D measurement. The blurring due to volumetric scattering makes it challenging to see the target directly from the measurement.} 
    \label{fig:setup}
\end{figure}

Imaging through dense, volumetric, scattering media has been a significant challenge for various applications, such as imaging through skin tissue and industrial remote sensing through the smoke. In conventional imaging, photons directly travel from the target to the sensor. However, in scattering media, photons randomly move, blurring the imaging target both in time and space (Figure~\ref{fig:setup}, right). This blur makes it challenging to resolve small features of the target from the measurement. 

Several approaches have been developed to tackle the challenge of imaging through a scattering medium. When the medium is not dense or thick, ballistic photons, that do not undergo scattering, can be used, effectively ignoring the effect of scattering. Optical coherence tomography~\cite{Huang91:OCT}, confocal microscopy~\cite{Webb96:Confocal}, and two-photon imaging~\cite{Helmchen2005DeepTT} optically reject scattered photons to capture only ballistic photons. Time-of-flight measurements enable the time-gating of ballistic photons~\cite{Wang:91}, so that the measurement only relies upon the first photon to hit the sensor. Methods to measure ballistic photons have been used to solve remote sensing problems in computer vision and robotics applications~\cite{Wang2018ProgrammableTL, Achar:2017:ETI,O'Toole:2012:PCP}. While these techniques can achieve cellular-level resolution when imaging through a thin medium, the number of ballistic photons decreases exponentially with the thickness of the scattering medium. For dense and thick media, the number of detected ballistic photons is below the noise level of the measurement, causing the reconstruction quality from early-arriving photons to degrade drastically.

Another approach is used by speckle-based techniques~\cite{Bertolotti12, Katz14}, which enable imaging through scattering media with diffraction-limited resolution. However, the memory effect~\cite{Freund88} limits the field-of-view for these techniques, which becomes impractically small when imaging in deep tissue. Speckle patterns can be modeled as a sum of complex fields. The reconstruction of a 2D object can be achieved by inverting a complex-valued linear system~\cite{Dremeau:15, Metzler17}, but capturing this linear system requires tedious calibrations with a fixed target and medium, making this method impractical for various applications.

When the number of scatterings through the medium is large, the propagation of photons can be modeled as diffusion. In the field of biomedical imaging, diffuse optical tomography (DOT)~\cite{Gibson05,Boas2001DOT} uses diffusion models of light through tissue and has seen substantial success. This method uses multiple light sources and detectors to generate a 3D reconstruction of the tissue being imaged and has been developed for few decades, with advancements including the use of time-of-flight measurement-based techniques~\cite{Kumar:06, Pifferi16:TDDOT, Lyons19:ToFDoT, Gkioulekas2016AnEO,Niedre06}. It has already been demonstrated in clinical applications~\cite{Corlu:07, Choe:09}.

In this paper, we exploit automatic differentiation for All Photons Imaging~\cite{Satat16:API} to recover an object inside volumetric scattering media. In contrast to traditional DOT techniques that require a set of illumination sources and detectors in contact with the scattering media, we achieve the reconstruction of the object shape with a simple transmission-mode imaging system (Fig.~\ref{fig:setup}). The difference between traditional DOT and All Photons Imaging is highlighted by Satat et al~\cite{Satat16:API}. The major advantages of API over traditional DOT are single-shot capability (DOT requires raster scanning), ease of calibration, and applicability to non-contact remote sensing. In this research, we capture the time-resolved measurement of the spatio-temporal profile of the scattered photons and recover the optical properties and 2D target embedded in the dense scattering media. While the diffusion approximation we use is well studied in the past~\cite{Niedre06, Patterson:89}, our adaptation of a modern automatic differentiation framework~\cite{jax} enables the robust estimation of the optical properties in complex scenes with a 2D target embedded within the scattering media. 

\begin{figure}
        \centering
        \includegraphics[width=0.7\linewidth]{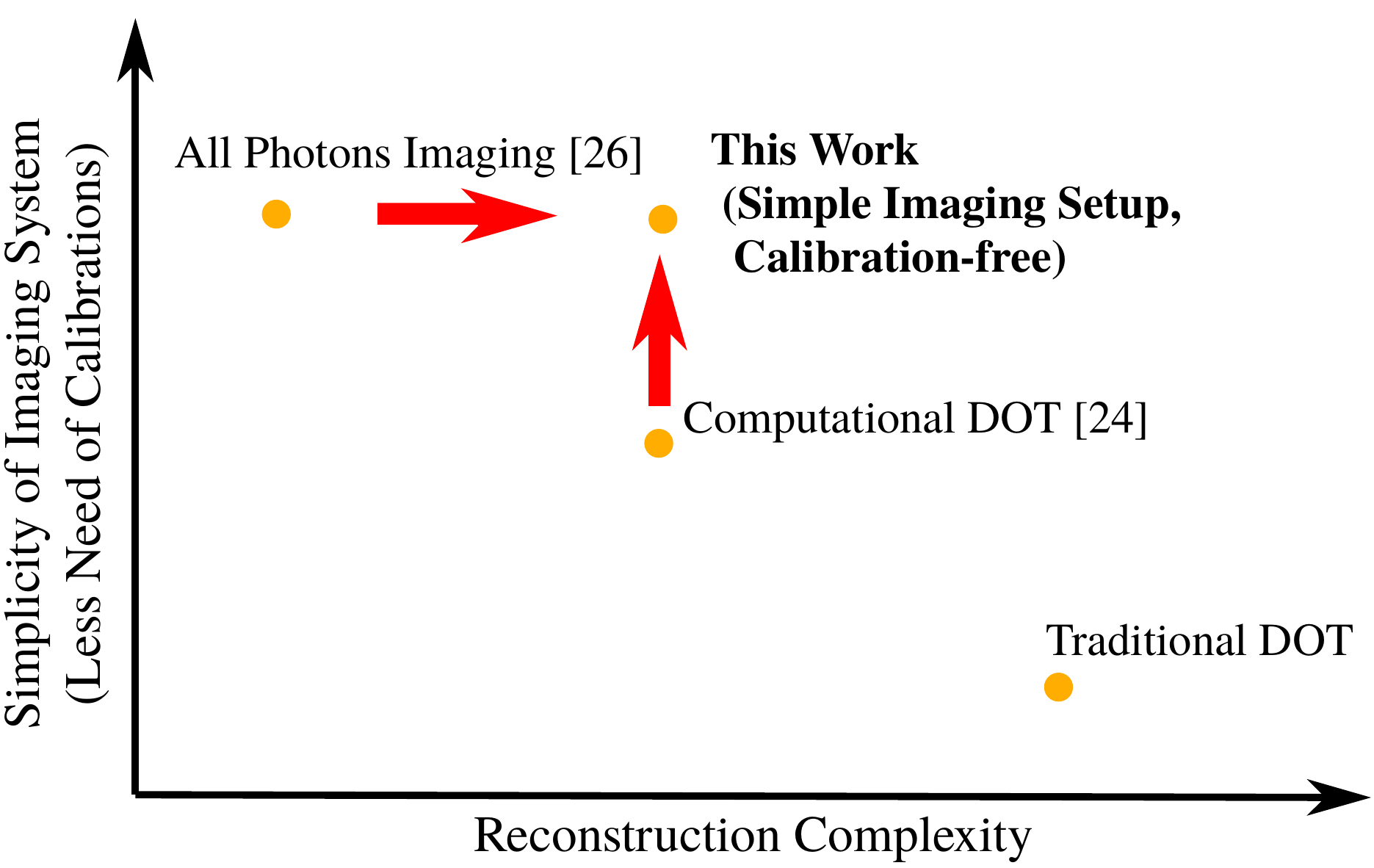}
        \caption{\textbf{Towards imaging more complex scene with simple imaging system}. Our technique exploits auto-differential diffusion approximation to simultaneously recover the target and the optical properties of the scattering media.} 
    \label{fig:motivation}
\end{figure}

\paragraph{Relation to closely related prior art:} Our method uses the same imaging system as All Photons Imaging (API)~\cite{Satat16:API}, but extends API to the a realistic scene (a 2D target embedded inside a phantom, instead of the target on top of a phantom) with the use of automatic differentiation to the forward model. We use the same optical setup as Lyons et al.~\cite{Lyons19:ToFDoT}. While Lyons et al.~\cite{Lyons19:ToFDoT} required a prior knowledge on the optical properties of the scattering media, our technique jointly recovers the unknown optical properties and embedded 2D target. In contrast to traditional DOT, our technique requires fewer priors, such the position of illumination sources and detectors, and is capable of single-shot measurements. Fig~\ref{fig:motivation} summarizes the relation of this work to previous works.
\section*{Results}
\subsection*{Forward Model}
The radiative transfer function (RTE) describes the random walk of photons through a homogeneous medium:
\begin{equation}
\resizebox{1.0\hsize}{!}{
    $\frac{1}{c} \frac{\partial L(\mathbf{r},\mathbf{\omega},t)}{\partial t} + \mathbf{\omega} \cdot \nabla L(\mathbf{r},\mathbf{\omega},t) = -\mu_t L(\mathbf{r},\mathbf{\omega},t) + \mu_s \int L(\mathbf{r},\mathbf{\omega},t) P(\omega, \omega^\prime) d\Omega' + S(\mathbf{r},\mathbf{\omega},t),$
    }
\end{equation}
The left two terms represent the change of the energy (per unit volume per unit solid angle) and the divergence of the beam, while the right three terms represent the extinction, scattered energy from the surrounding, and light source. $\mu_a$ and $\mu_s$ respectively represent the expected number of of absorption and scattering per unit light path, and $\mu_t$ is a sum of $\mu_a$ and $\mu_s$.  The terms $\mathbf{r},\mathbf{\omega},t$ denote the location, direction and time respectively. $c$ is a speed of light in the media. $L(\mathbf{r},\mathbf{\omega},t)$, $ P(\omega, \omega^\prime)$ denote the radiance and phase function respectively. The phase function is often represented by the anisotropy parameter $g = <\cos\theta>,$ where $\theta$ is the angle between the old and new trajectories.

The RTE is hard to solve, so two approaches are often used to approximate. When the number of scatterings is sufficiently large and $\mu_a << \mu_s$, the diffusion equation approximates the RTE, where the photon fluence rate $\phi(\mathbf{r},t) = \int L(\mathbf{r},\mathbf{\omega},t) d\Omega$ satisfies the diffusion equation. The diffusion approximation becomes less accurate as the number of scatterings becomes small, as the propagation of photons cannot be considered diffusion. In situations with limited scattering, a Monte Carlo approximation is used to approximate the RTE. However, Monte Carlo approximations are computationally expensive as they involve the simulation of the random walks of millions of photons.

Here, we consider dense and thick scattering media, where the diffusion approximation can provide a reasonably accurate solution to the RTE. The diffusion equation is 

\begin{equation}
    \frac{1}{c}\frac{\partial}{\partial t}\phi(\mathbf{r},t) - D\nabla^2\phi(\mathbf{r},t)  + \mu_a \phi(\mathbf{r},t) = S(\mathbf{r},t),
\label{eq:diffusion_equation}
\end{equation}
where $D = [3(\mu_a + \mu_s(1-g))]^{-1}$ is the diffusion coefficient. For an isotropic light source with short time pulse, the photon fluence rate in the infinite scattering media is written as
\begin{equation}
    \phi(\mathbf{r},t) = c(4\pi Dct)^{-3/2}\exp\bigg(-\frac{r^2}{4Dct} - \mu_a ct\bigg).
\end{equation}.

\begin{figure}
        \centering
        \includegraphics[width=1.0\linewidth]{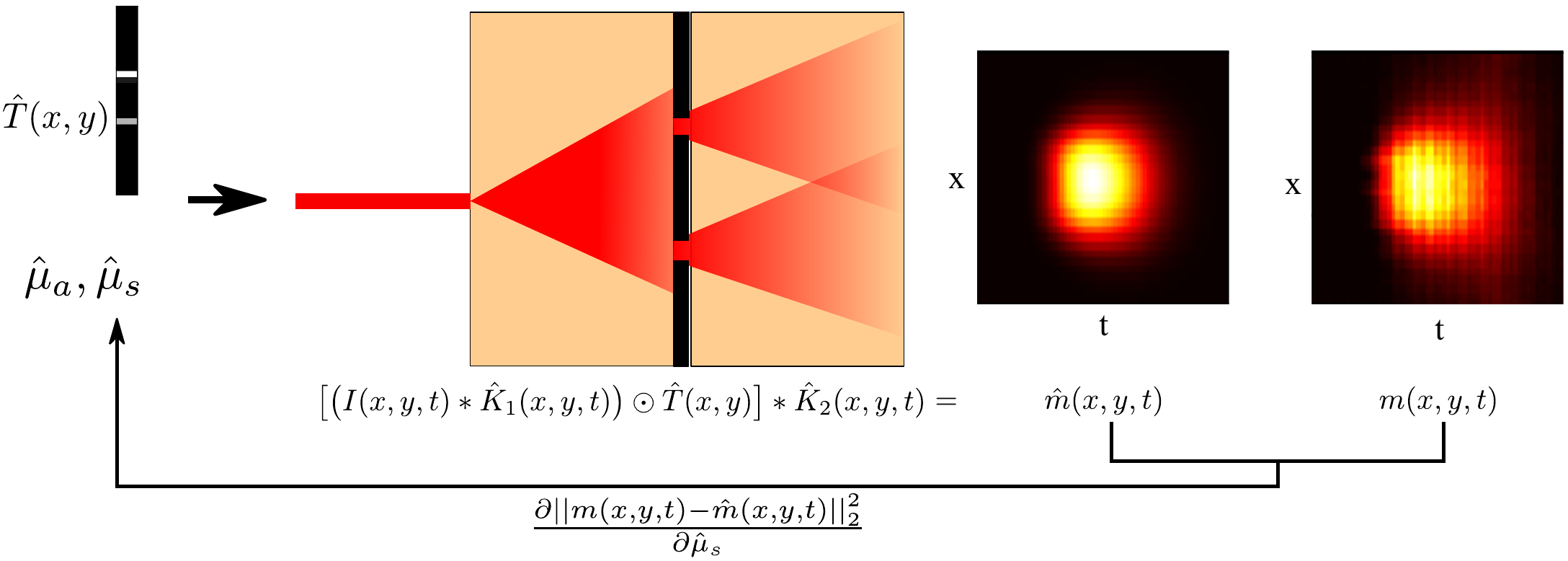}
        \caption{\textbf{A sketch of the forward model and reconstruction.} (a) The illumination ($I(x,y)$) is blurred as photons scatter in the first scattering media ($K_1(x,y,t; D, d_1)$) before passing through the 2D target ($T(x,y)$). Photons scatter in the second scattering media ($K_2(x,y,t; D, d_2)$) before reaching the detector. We illustrate the forward model in 2D for simplicity. Our reconstruction method takes the 3D measurement over x, y, and t for input. (b) Our technique estimates the unknown optical properties of the media by minimizing the observed measurement $m(x,y,t)$ and estimated measurement ($\hat{m}(x,y,t)$).} 
    \label{fig:overview}
\end{figure}

As illustrated in Fig.~\ref{fig:overview}, our imaging setup is in transmission mode, where a target is between scattering media. Unlike traditional DOT, the goal of our technique is to reconstruction a 2D target within a scattering media with high contrast. We express the measurement as the result of a 3D convolution with $K_1(x,y,t)$ (scattering between illumination and target), followed by a 2D multiplication with $T(x,y)$ (2D mask) and another 3D convolution with $K_2(x,y,t)$ (scattering between target and detector):
\begin{equation}
    m(x,y,t) = [\{I(x,y,t) * K_1(x,y,t; D,d_1)\} \odot T(x,y)] * K_2(x,y,t; D, d_2).
\label{eq:measurement}
\end{equation}

$I(x,y,t)$ denotes the illumination on the surface of the scattering media. Our forward model is similar to the model proposed by Lyons et al.~\cite{Lyons19:ToFDoT}. In this paper, we consider a 2D target with sparse opening pixels. Therefore, the geometry of the phantoms before and after the phantom can be approximated as slobs. We follow diffusion approximation model proposed by Patterson et al.~\cite{Patterson:89} to write the 3D kernel $K_1(x,y,t; D, d_1), K_2(x,y,t; D, d_2)$ that represent the propagation of photons through the scattering media with thickness $d$ as follows.
\begin{equation}
\resizebox{0.95\hsize}{!}{$
\begin{split}
K(x,y,t; D,d) &= (4 \pi Dc)^{-3/2}t^{-5/2}\exp \left(-\mu_a ct \right)\exp\left(-\frac{|r|^2}{4Dct}\right)  \\ & \times \bigg[(d-z_0)\exp \left(-\frac{(d-z_0)^2}{4Dct}\right) - (d+z_0)\exp \left(-\frac{(d+z_0)^2}{4Dct} \right) \\ & + (3d-z_0)\exp \left(-\frac{(3d-z_0)^2}{4Dct} \right) - (3d+z_0)\exp \left(-\frac{(3d+z_0)^2}{4Dct}\right)\bigg],
\end{split}
$}
\label{eq:slab_dissusion}
\end{equation}
where $z_0 = [\mu_s(1-g)]^{-1}$ and $|r|^2 = x^2 + y^2 + d^2$. This is different from Eq.~\ref{eq:diffusion_equation} because we consider the photon transmittance of a homogeneous slab media that is bounded along optical axis, and infinite in the other two axes. Applying two boundary conditions for surfaces of the slab geometry results in Eq.~\ref{eq:slab_dissusion}.

\subsection*{Reconstruction Algorithm}
Our method estimates the 2D target as well as the optical properties of the homogeneous scattering media from the measurement. We write the objective of the algorithm as follows. 

\begin{equation}
\begin{split}
    \min \norm{m(x,y,t) - [\{I(x,y,t) * K_1(x,y,t; D,d_1)\} \odot T(x,y)] * K_2(x,y,t; D,d_2)}^2 \\ + \Gamma(T(x,y)),
\end{split}
\label{eq:loss}
\end{equation}
where $\Gamma(T(x,y))$ is a regularization on the target. We solve this problem in the model by minimizing with respect to the optical property and the target in an alternate manner.

Scattering coefficient $\mu_s$, anisotropy g, and absorption coefficient $\mu_a$ are commonly used to represent the scattering properties of media. In the diffusion paradigm, the scattering coefficient and anisotropy are interchangeable, so reduced scattering coefficient $\mu_s^\prime = \mu_s(1-g)$ is used instead. While we assume $\mu_a << \mu_s$, $\mu_a$ still has a significant impact on the 3D kernel $K(x,y,t)$. Hence, our goal is to estimate $\mu_s^\prime$ and $\mu_a$ from the measurement. 

\paragraph{Estimation of $\mbox{\boldmath$\mu_a$}$:}
As Chance et al.~\cite{Chance1988TimeresolvedSO} suggest, the absorption coefficient $\mu_a$ can be estimated with the asymptotic slope of the log of the measurement along the time dimension. Let us consider the blur kernel $K(x,y,t; D,d)$ for large $t$ (See Supplemental Material for the derivation):
\begin{equation}
\begin{split}
    \lim_{t=\infty}\frac{\partial}{\partial t}\log{K(x,y,t;D, d)} = \mu_a c.
\end{split}
\label{eq:mua_limit}
\end{equation}
Hence. $t$ is sufficiently large $\mu_a$ can be estimated as 
\begin{equation}
    {\hat{\mu}_a} \approx -\frac{1}{c}\frac{\partial}{\partial t}\log{K(x,y,t=t_l; D, d)}.
\label{eq:mua_estimate}
\end{equation}
Our forward model is more complex than the expression of the 3D kernel in Eq.~\ref{eq:slab_dissusion}. However, the insight from Eq.~\ref{eq:mua_limit} shows that the slope of the log of the measurement is dominantly determined by $\mu_a$. Hence, we estimate the absorption coefficient as written in Eq.~\ref{eq:mua_estimate}.

\paragraph{Estimation of \mbox{\boldmath$\mu_s^\prime$}:}
While $\mu_a$ can be estimated because absorption dominates the intensity fall-off for sufficiently large $t$, an analytical expression of $\mu_s^\prime$ in terms of the measurement is challenging. We numerically estimate $\mu_s^\prime$ with gradient descent. As shown in Eq.~\ref{eq:measurement} and Eq.~\ref{eq:slab_dissusion}, the forward model is fairly complex. Our work uses recent advancements in auto-differentiation to calculate the gradient of the objective function (Eq.~\ref{eq:loss}) with respect to $\mu_s^\prime$. Automatic differentiation efficiently and accurately computes the derivative of functions from the forward model. It has been exploited for robust design optimization~\cite{renaud1997automatic} computational fluid dynamics~\cite{Bischof92AD}, and has recently gained significant interest from machine learning community~\cite{Baydin:2017}. In imaging, automatic differentiation has been used in DOT with more complex forward model~\cite{Arridge:98,Klose02:ADDOT}, fluorescent lifetime imaging~\cite{Roy:99}. We use JAX~\cite{jax} to implement the forward model in Eq.~\ref{eq:measurement} and compute the gradient with respect to $\mu_s^\prime$, and RMSProp as the optimizer~\cite{hinton2012neural}. 
\\ While Patterson et al.~\cite{Patterson:89} suggest a closed-form estimate of the reduced scattering coefficient from the measurement on slab geometry, our imaging setup contains a 2D target in the media, which introduces model mismatch. Moreover, such a closed-form solution depends heavily on the estimate of the absorption coefficient, which can vary in low SNR regime. Therefore, a reduced scattering coefficient estimated by this method is not accurate, resulting in an inferior reconstruction of the target (See Supplemental Material Section 3). The use of automatic differentiation on the forward model with diffusion approximation enables robust recovery of the target, as our technique aims to minimize the loss between the expected measurement and the actual measurement. 

\paragraph{Estimation of the target:}
Given $\mu_s^\prime$ and $\mu_a$, the measurement can be computed by Eq.~\ref{eq:measurement}. The target can be recovered by gradient descent with the gradient computed from automatic differentiation. However, the model can be simplified as a linear system to solve the minimization problem efficiently: 
\begin{equation}
    \mathbf{b} = \mathbf{Av},
\label{eq:lineareq}
\end{equation}
where $\mathbf{b}$ and $\mathbf{v}$ denote the vectorized 3D measurement $m(x,y,t)$ and 2D target $T(x,y)$ respectively. $\mathbf{A}$ represents the mapping between the measurement target, and can be obtained by computing the expected measurement for each pixel of the target. For example, $i$th column of $\mathbf{A}$ is the expected measurement for the case $\mathbf{v}_i = 1$, and given as
\begin{equation}
    \mathbf{A}_i = \text{vec}\bigg([\{I(x,y,t) * K_1(x,y,t; D, d_1)\} \odot T_i(x,y)] * K_2(x,y,t, D, d_2)\bigg),
\end{equation}
where $T_i(x,y)$ is 1 for the $i$th pixel and 0 for all the other pixels. vec($\cdot$) denotes an operator that vectorizes 3D matrix into a normalized column vector. Eq.~\ref{eq:lineareq} is ill-posed because $\mathbf{A}$ is a blurring operator. Therefore, we solve this problem with l1 and total variation regularizer:
\begin{equation}
    \hat{\mathbf{v}} = \argmin_\mathbf{v} \frac{1}{2} \norm{\mathbf{b} - \mathbf{Av}}^2+ \Gamma(\mathbf{v}).
\label{eq:regularized_loss}
\end{equation}
The prior term is written as follows:
\begin{equation}
\begin{split}
\Gamma(\mathbf{v}) &= \lambda_1 \norm{v}_1 + \lambda_2 \norm{\nabla v}_1 \\ &= \norm{\mathbf{Gv}}_1,
\end{split}
\end{equation}
with 
\begin{equation}
    \mathbf{G} = \begin{bmatrix} \mathbf{I} \\ \frac{\lambda_2}{\lambda_1}\mathbf{D_x} \\ \frac{\lambda_2}{\lambda_1}\mathbf{D_y} \end{bmatrix},
\end{equation}
where $D_x, D_y$ are matrices that compute the gradient of the target image in x and y directions, respectively.

We use the alternating direction method of multipliers (ADMM)~\cite{Boyd:2011:ADMM} to solve this minimization problem. The iterative steps of ADMM are summarized in Alg.~\ref{alg:ADMM_Algorithm}. 

While we use l1 sparsity prior to our target, our formulation could extend to types of common priors such as total-variation or sparsity over the discrete some bases (e.g., discrete cosine transform). When such priors are not available, Tikhonov (l2) regularization can be used (See Supplemental Material).

\begin{algorithm}[t]
\KwIn{Input: $\mathbf{v}^0, \mathbf{z}^0 = \text{vec}(\hat{T}(x,y)), \quad \mathbf{u}^0 = 0, \quad \rho,\lambda_1, \lambda_2$}
\While {not converged}{
\nl $\mathbf{v}^{i+1} = (\mathbf{A^\top A} + \rho \mathbf{G}^\top \mathbf{G})^{-1} (\mathbf{A^\top b} + \rho \mathbf{z}^i - \mathbf{u}^i).$ \\
\nl $\mathbf{z}^{i+1} = \max\big(\lvert \mathbf{v}^{i+1} - \mathbf{u}^i  \rvert - \lambda / \rho, 0).$ \\
\nl $\mathbf{u}^{i+1} = \mathbf{u}^i + \mathbf{v}^{i+1} - \mathbf{z}^{i+1}.$
}
\label{alg:ADMM_Algorithm}
\caption{\textbf{ADMM algorithm to solve Eq.~\ref{eq:regularized_loss}}}
\end{algorithm}

\subsection*{Experimental Demonstration}
We demonstrate our technique with both simulated and experimental data. 

\begin{figure}
        \centering
        \includegraphics[width=1.0\linewidth]{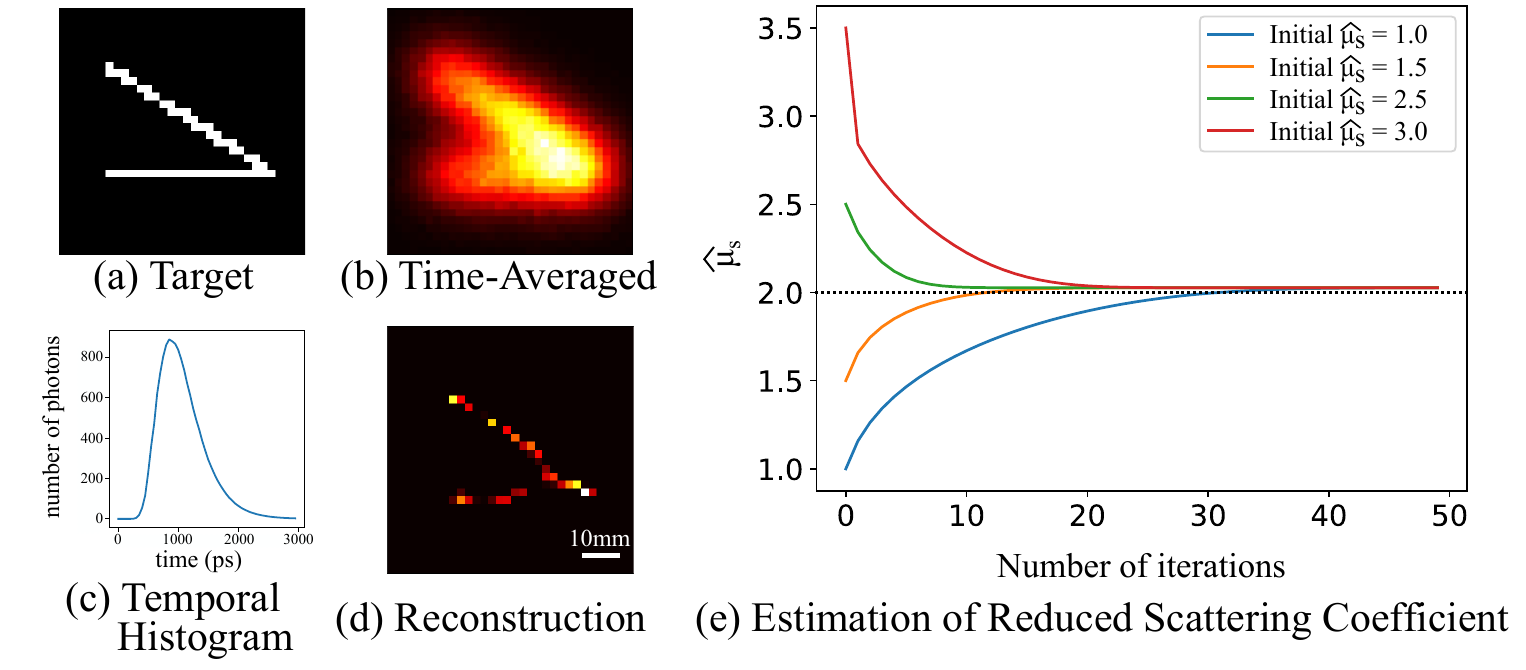}
        \caption{\textbf{Reconstruction results on the measurement simulated by Monte Carlo renderer.} (a) A Wedge-shaped target was used for rendering. (b), (c) The measurement is blurred in space and time due to scattering. (d), (e) Our technique recovers the target and the optical property of the scattering media. (e) Different initialization of the reduced scattering media converges close to the ground truth.} 
    \label{fig:simulation_results}
\end{figure}

\paragraph{Simulation:}
We use a Monte Carlo renderer to simulate the time-of-flight measurements in our setup and validate our estimation of optical properties and the target. The simulated measurement does not contain any sensor noise, such as dark noise. The scattering coefficient, anisotropy, and absorption coefficient of the simulated phantom were 20.0 mm\textsuperscript{-1}, 0.9, and 0.01 mm\textsuperscript{-1}. The reduced scattering coefficient is 2.0 mm\textsuperscript{-1}. Fig.~\ref{fig:simulation_results} shows the recovered target and reduced scattering coefficient for the first 50 iterations of $\mu_s^\prime$ estimation. The estimated reduced scattering coefficient and the absorption coefficient was 2.04 mm\textsuperscript{-1} and 0.011 mm\textsuperscript{-1}. Fig.~\ref{fig:simulation_results} shows that different initialization converges close to the ground truth.

\begin{figure}
        \centering
        \includegraphics[width=1.0\linewidth]{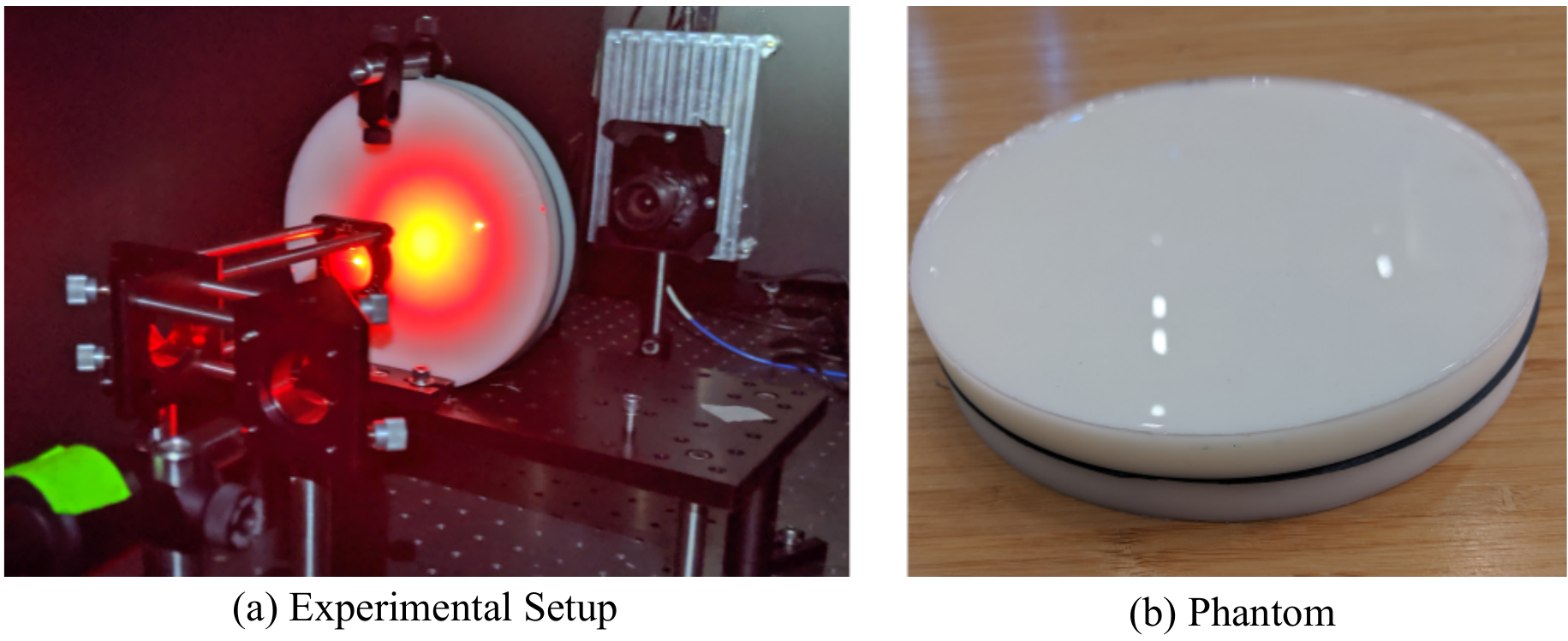}
        \caption{\textbf{Experimental setup and the scattering media.} (a) Picture of the transmission-mode imaging system. (b) The phantom is a dense volumetric media. The object is in the middle of the phantom.} 
    \label{fig:experiment_picture}
\end{figure}

\begin{figure}
        \centering
        \includegraphics[width=1.0\linewidth]{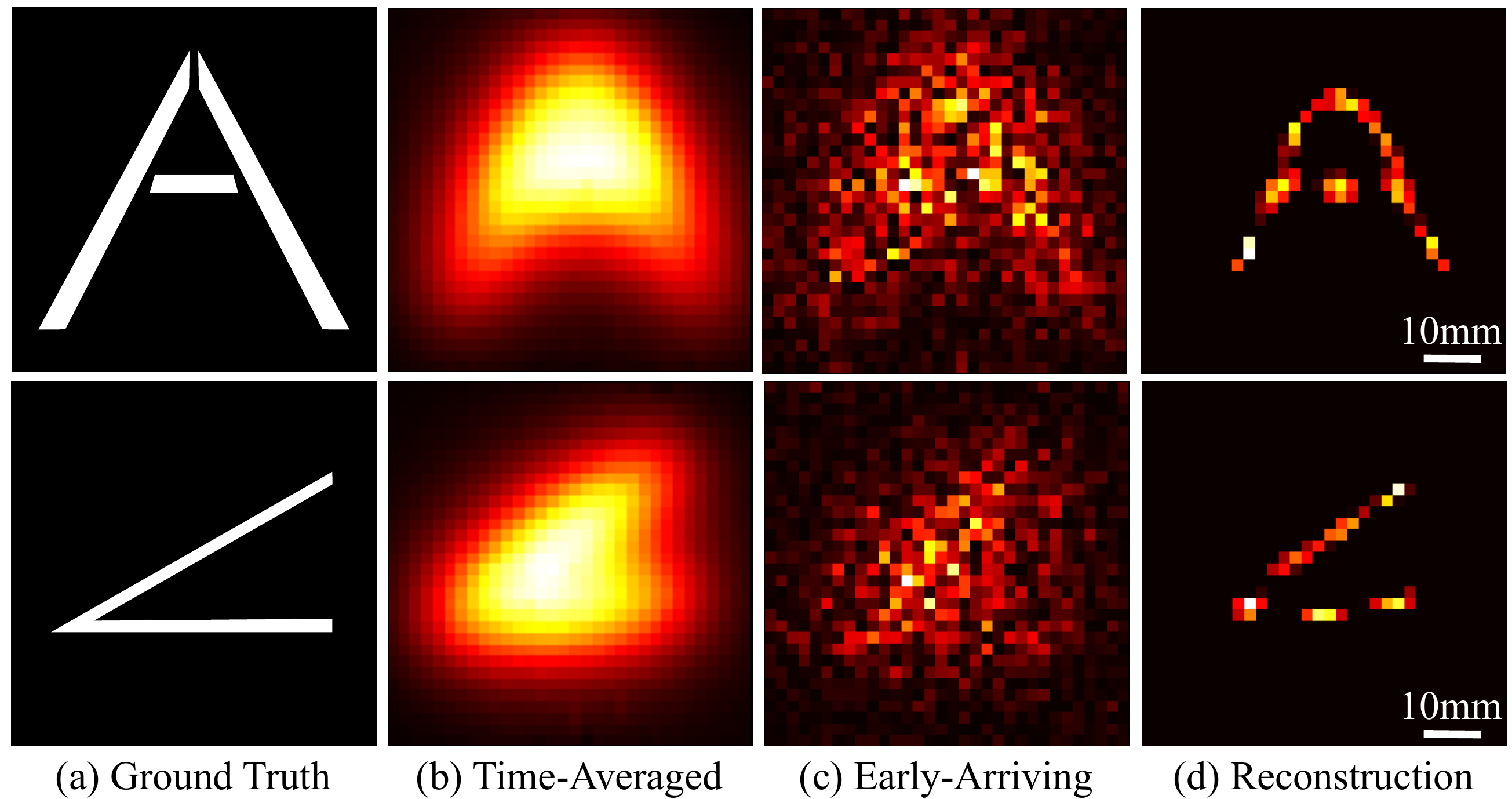}
        \caption{\textbf{Reconstruction results on the measurement from an experiment with 2D targets.} (a) Ground Truth. (b) Measurement integrated over time. (c) A frame with first arriving photons are captured. (d) Reconstruction.} 
    \label{fig:experiment_results}
\end{figure}

\paragraph{Experiment:} 
Fig.~\ref{fig:experiment_picture} illustrates the experimental setup. The thickness of the phantom was 26 mm in total, where the target is placed in the middle. Fig.~\ref{fig:experiment_results} shows the reconstruction results in 32 by 32 resolution with a pixel size of 2 mm, compared with time-averaged and ballistic-photon imaging. We integrated 2D frames with respect time for the time-averaged frame, which is equivalent to an image taken without the temporal resolution. To generate a frame of early-arriving photons, we extract the first photons above the noise level for each pixel. (See supplemental video for the raw 3D measurement.) The time-averaged frame shows blur due to the scattering, and the ballistic frame is noisy because the number of ballistic photons is small. Our method successfully recovers the shape of the 2D target. The estimated reduced scattering and absorption coefficient are 2.45 mm\textsuperscript{-1} and 0.0082 mm\textsuperscript{-1} for "A" target, and 2.39 mm\textsuperscript{-1} and 0.0088 mm\textsuperscript{-1} for the "wedge" target.

\begin{figure}
        \centering
        \includegraphics[width=0.6\linewidth]{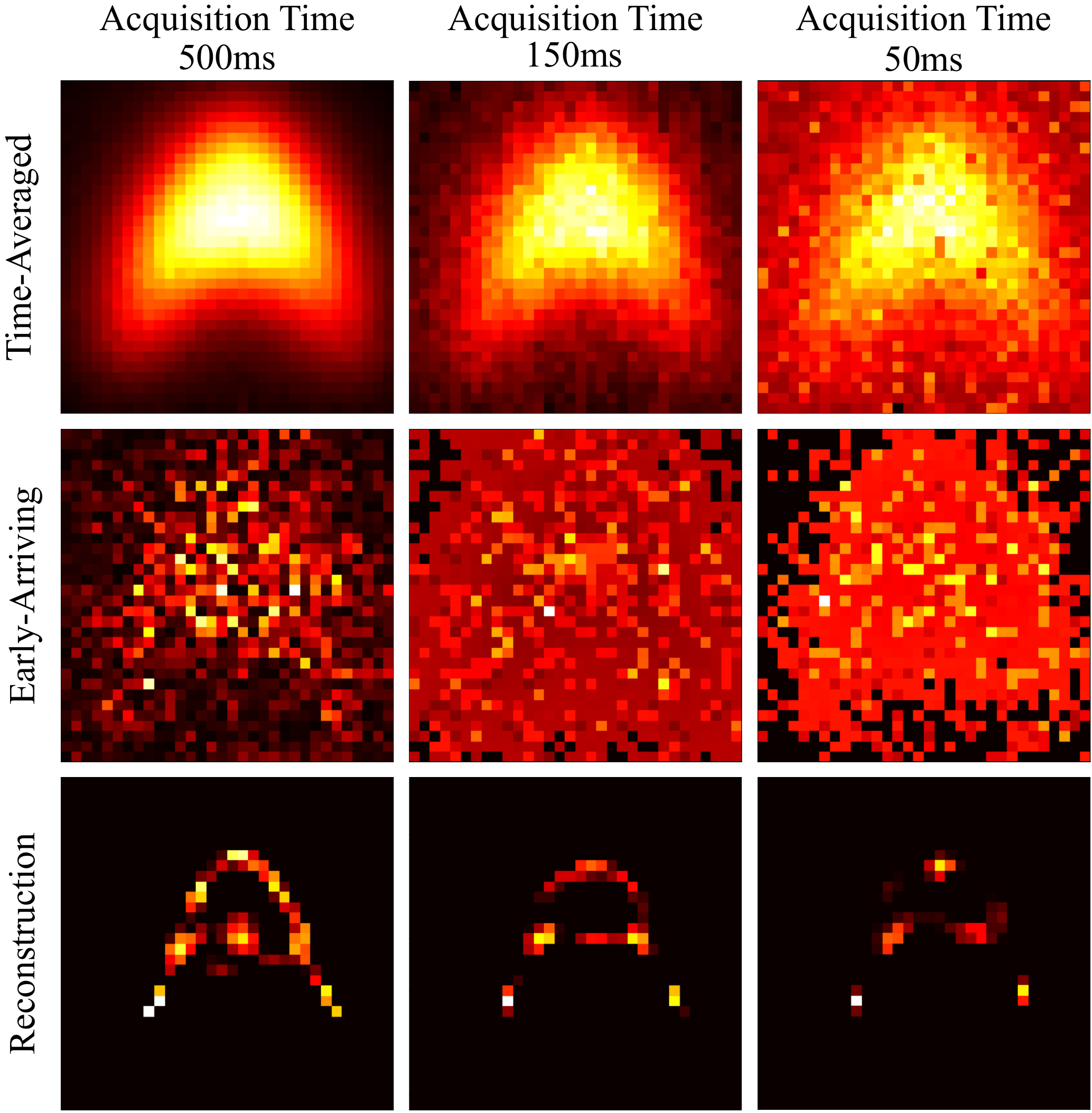}
        \caption{\textbf{Reconstruction with different acquisition times.} Our technique recovers the target with 150 ms acquisition time, showing the benefit of simple imaging setup with a single-shot capability.} 
    \label{fig:compare_frame_nums}
\end{figure}

\begin{figure}
        \centering
        \includegraphics[width=0.9\linewidth]{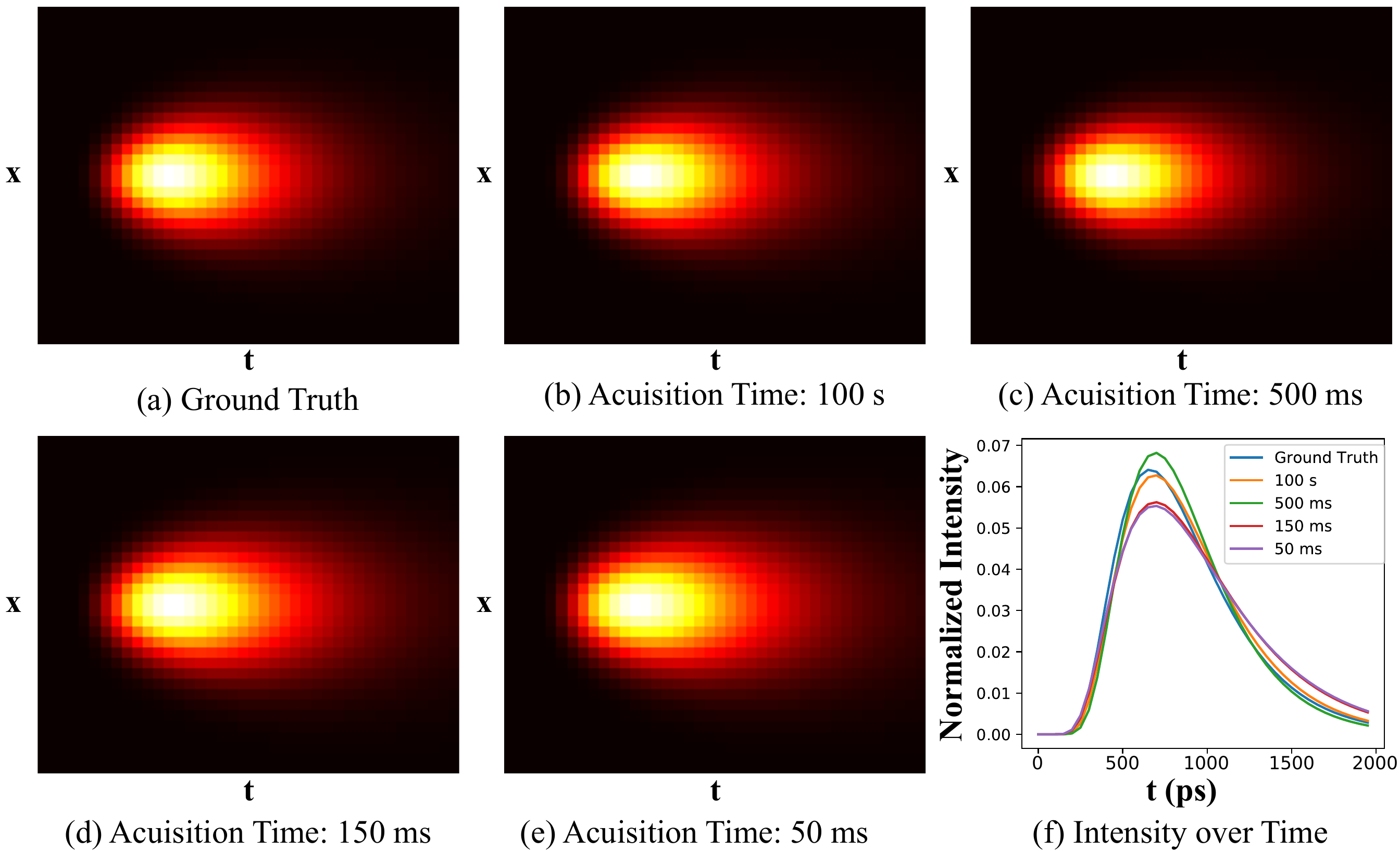}
        \caption{\textbf{Estimated blur kernels for measurements with different acquisition times.} The accuracy of the optical properties becomes worse with noise due to a shorter acquisition time. However, our technique is still able to estimate a blur kernel, which is close to the ground truth. (a)--(e) shows the x-t plot of the intensity, and (f) shows the plot of normalized intensity over time for the estimates of the blur kernel with different acquisition times.} 
    \label{fig:kernel_estimation}
\end{figure}

Fig.~\ref{fig:compare_frame_nums} shows reconstruction results for different acquisition time. Our method is robust even when the SNR of the measurement is so low that the time-averaged and ballistic frames are noisy. Our technique reconstructed the "A" target with 150 ms exposure time. This shows the benefit of our simple optical setup. Fast acquisition is challenging with traditional DOT, which requires raster scanning of the illumination source. The accuracy of the estimation of the optical properties decreases as the acquisition time decreases. For example, for the acquisition time of 50ms, the estimated reduced scattering coefficient and absorption coefficient are 2.04 mm\textsuperscript{-1} and 0.0031 mm\textsuperscript{-1}. This is due to the estimation of the absorption coefficient from the noisy temporal histogram from a small number of photons. However, as shown in Fig.~\ref{fig:kernel_estimation} (a)--(e), the estimated blur kernels $K(x,y,t;D,d)$ for different acquisition times are still similar to the ground truth. Fig.~\ref{fig:kernel_estimation} (f) plots the normalized intensity over time. The estimated kernel roughly follows the ground truth blur kernel, which we generated from the estimate of the optical properties with phantoms with no target.

\section*{Discussion}

\paragraph{Practical limitations:}
In this paper, we demonstrate reconstruction of a 2D mask in an unknown homogeneous volumetric scattering media. Our technique extends the limitations of previous works in complexity~\cite{Satat16:API} or ease of use without calibration~\cite{Lyons19:ToFDoT}. The use of automatic differentiation provides a solution towards practical use of previously proposed works. However, our technique still requires prior knowledge of the target depth, and the transmission-mode imaging system limits the reconstruction to 2D shapes. We also only consider flat surfaces for the illumination and detection.  

\begin{figure}
        \centering
        \includegraphics[width=0.9\linewidth]{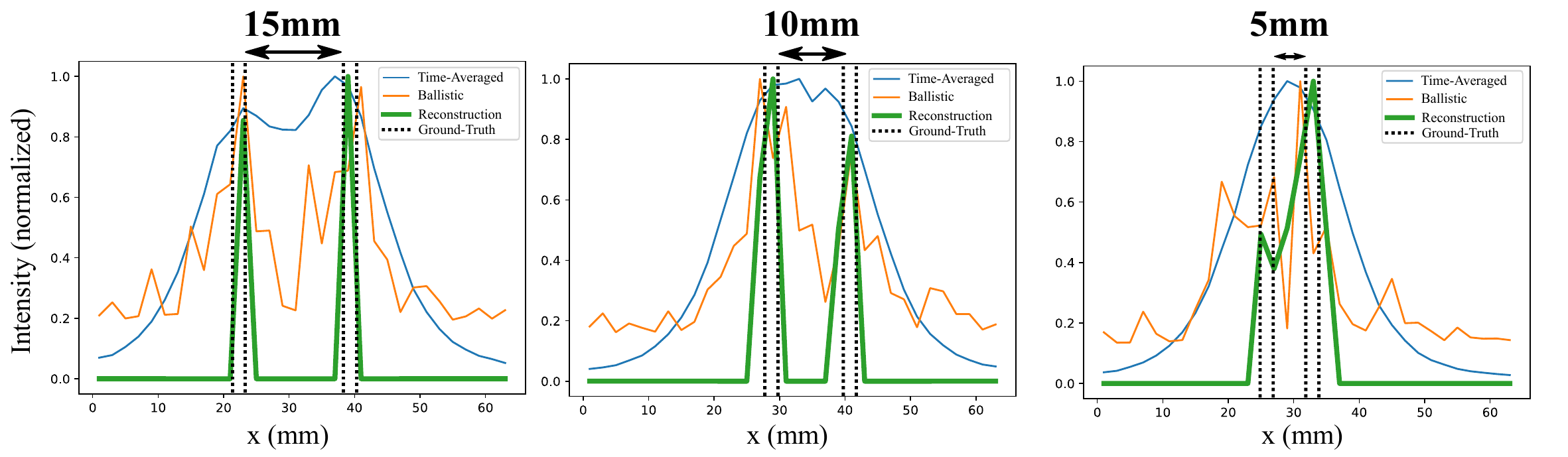}
        \caption{\textbf{Reconstruction of bars separated by 15 mm, 10 mm, 5 mm space to evaluate the resolution of our methods compared to time-averaged and ballistic reconstruction.} Time-averaged reconstruction suffers from blur due to scattering, and ballistic reconstruction suffers from low SNR. Our method resolves two points separated by 15mm and 10mm.} 
    \label{fig:resolution}
\end{figure}

\paragraph{Resolution:}
We used 2D targets that consist of two 2 mm by 10 mm bars that are 5 mm, 10 mm, 15 mm away from each other to evaluate the resolution of our system. Figure~\ref{fig:resolution} shows the 1D reconstruction of the targets. This illustrates the resolution of our method for the same scattering media described in the experimental demonstration as compared to time-averaged and ballistic photon imaging. The resolution of imaging is determined by the blur from the scattering media. The illustrated resolution should be worse as the scattering coefficient or the thickness of the phantom increase. Traditional DOT places illumination sources and detectors at different locations to capture more information for better reconstruction quality. In contrast, the goal of our technique is to improve the practicality of the imaging system. 

\paragraph{Need for Time-of-Flight Measurement:}
Time-of-flight information plays an important role in the estimation of the diffusion coefficient. Our algorithm initializes an estimation of the target with the time-averaged frame. Without the time of flight, the estimation of the blur kernels $K_1(x,y,t; D, d_1)$ and $K_2(x,y,t; D, d_2)$ becomes an impulse response as the estimate becomes equivalent to the measurement.

\paragraph{The Use of Diffusion Approximation and Linear Model:}
While the linear model provides an approximation to RTE, the accuracy of the reconstruction is limited. Furthermore, the linear model may introduce an error because it does not model photon's random walk from the second phantom back to the first phantom across the target plane. Therefore, Monte Carlo-based reconstruction should produce more accurate reconstruction result~\cite{Gkioulekas2016AnEO}. However, such a method requires many iterations of Monte-Carlo rendering, which is computationally expensive. In contrast, diffusion-approximation-based reconstruction is more computationally efficient. In practice, our method can be used for the initialization of the Monte Carlo-based approach for faster convergence.

\paragraph{Conclusion:}
We propose a novel algorithm to reconstruct a mask in dense, scattering media with unknown optical properties and demonstrate that it can reconstruct targets embedded in thick, scattering media with limited calibration. The automatic differentiation to the forward model overcomes the challenge of All Photons Imaging, and enables recovery of object embedded in the dense volumetric scattering media. We hope that our approach inspires more practical imaging through scattering media without extensive calibrations for biomedical and remote sensing applications. 

\section*{Methods}
\paragraph{Optical setup:}
Fig.~\ref{fig:experiment_picture} describes the optical setup. The SPAD array is focused on the surface of the tissue phantom, and one pixel corresponds to a 2 mm square patch on the surface. The illumination source is a SuperK super-continuum laser, where white light is filtered at 628 nm with acousto-optic tunable filters followed by a 50-degree circle diffuser (ThorLab, ED1-C50-MD), which was placed 6.4cm away from the phantom. The instrument response has roughly 150 ps full-width-half-max (FWHM). Each measurement is 32 x 32 x 70 tensor with a 50ps time bin. Exposure time is 0.5 ms for each frame of the SPAD array, and we measure 200000 frames for the complete measurement. The total acquisition time to generate reconstruction in Fig.~\ref{fig:experiment_results} is 100 seconds. The biomimetic phantom consists of epoxy resin (East Coast Resin) as a base and TiO\textsubscript{2} powder (Pure Organic Ingredients) as a scattering agent~\cite{Diep:15}. The refractive index of the media is 1.4~\cite{Diep:15} and the total thickness of the phantom is 26mm (13mm on each side of the target). We measured the absorption coefficient of the phantom as 0.080 mm\textsuperscript{-1} using Eq.~\ref{eq:mua_estimate} and the reduced scattering coefficient as 2.32 mm\textsuperscript{-1} by fitting the measurement with Eq.~\ref{eq:slab_dissusion}. For simulated measurement, we use the same parameters for the detector, and the illumination source was placed 10 cm away from the phantom. Reconstruction was performed on a 32 x 32 x 60 3D x-y-t measurement with a 50 ps time bin.

\paragraph{Algorithm parameters:}
All the reconstruction on experimental data was performed with $\rho = 1$, $\lambda_1 = 8\times10^{-4}$, and $\lambda_2 = 10^{-3}\lambda_1$. The update of $\mu_s^\prime$ was performed with a step size of $5\times 10^{-2}$, and terminated when a certain number of iterations are performed or when the change of value is less than $0.01\%$ in a single iteration. Each $\mu_s$ and $T(x,y)$ has 300 and 50 iterations, alternated three times. The initial estimation of the reduced scattering coefficient was $1.5mm^{-1}$ for all of our reconstruction. We used the time-averaged frame as the initialization of the target estimate. The reconstruction on simulated data was performed with $\rho = 1$, $\lambda_1 = 1\times10^{-4}$, and $\lambda_2 = 10^{-3}\lambda_1$. Each $\mu_s$ and $T(x,y)$ has 300 and 50 iterations, alternated three times. We used larger $\lambda_1$ for the experimental data because of higher noise level.

\section*{Acknowledgments} 
We thank Prof. Moungi Bawendi, Nathan Klein and Collin Perkinson for their generous support of experimental hardware. Tomohiro Maeda and Ramesh Raskar are supported by Media Lab consortium funding and NSF Expeditions award IIS-1729931 and DARPA Reveal Program N00014-18-1-2894. Ankit Ranjan is supported by St John's College, Oxford Special Grant Fund. We reused materials such as figures from the author's thesis~\cite{Maedamastersthesis}.

\bibliographystyle{unsrt}  
\bibliography{references}  
\pagebreak

\section*{Supplemental Material}
\subsection*{Supplementary Videos}
Video S1 shows the 3D measurement of the "A" target for different exposure time (100 s, 500 ms, 150 ms, 50 ms) that are used for the reconstruction. Video S2 shows the 3D measurement for the "wedge" target. Video S3 shows the3D measurement for the bars that we used to evaluate the resolution of our imaging system.

\subsection*{Estimation of Absorption Coefficient} 
The left side of Eq.7 in the main manuscript can be expanded as follows:
\begin{equation}
\begin{split}
&\lim_{t=\infty} \frac{\partial}{\partial t} \log{K(x,y,t;D,d)} \\ & \quad= -\frac{5}{2t} - \mu_a c +  \frac{|r|^2}{4Dct^2} \\
& \quad + \frac{\partial}{\partial t} \log \bigg[ (d-z_0)\exp\bigg(-\frac{(d-z_0)^2}{4Dct}\bigg) - (d+z_0)\exp\bigg(-\frac{(d+z_0)^2}{4Dct}\bigg) \\ &\qquad + (3d-z_0)\exp\bigg(-\frac{(3d-z_0)^2}{4Dct}\bigg) - (3d+z_0)\exp\bigg(-\frac{(3d+z_0)^2}{4Dct}\bigg) \bigg].
\end{split}
\label{eq:expanded}
\end{equation}
We define $f(t)$ as following:
\begin{equation}
    \begin{split}
         f(t)&= (d-z_0)\exp\bigg(-\frac{(d-z_0)^2}{4Dct}\bigg) - (d+z_0)\exp\bigg(-\frac{(d+z_0)^2}{4Dct}\bigg) \\ &\quad+ (3d-z_0)\exp\bigg(-\frac{(3d-z_0)^2}{4Dct}\bigg) - (3d+z_0)\exp\bigg(-\frac{(3d+z_0)^2}{4Dct}\bigg). 
    \end{split}
\end{equation}
Since $\log(f(t))^\prime = f^\prime(t) / f(t)$ and 
\begin{equation}
\begin{split}
       \frac{\partial f(t)}{\partial t} &= \frac{(d-z_0)^3}{4Dct^2}\exp\bigg(-\frac{(d-z_0)^2}{4Dct}\bigg) - \frac{(d+z_0)^3}{4Dct^2}\exp\bigg(-\frac{(d+z_0)^2}{4Dct}\bigg)\\ & \quad+ \frac{(3d-z_0)^3}{4Dct^2}\exp\bigg(-\frac{(3d-z_0)^2}{4Dct}\bigg) - \frac{(3d+z_0)^3}{4Dct^2}\exp\bigg(-\frac{(d+z_0)^2}{4Dct}\bigg) \\ &=
    \frac{1}{4Dct^2} \bigg[ (d-z_0)^3 \exp\bigg(-\frac{(d-z_0)^2}{4Dct}\bigg) -   (d+z_0)^3 \exp\bigg(-\frac{(d+z_0)^2}{4Dct}\bigg) \\ &\quad + (3d-z_0)^3 \exp\bigg(-\frac{(3d-z_0)^2}{4Dct} - (3d+z_0)^3 \exp\bigg(-\frac{(3d+z_0)^2}{4Dct}\bigg)\bigg) \bigg].
\end{split}
\end{equation}
Since $f'(t)$ has $\frac{1}{t^2}$ term while $f(x)$ doesn't, $\lim_{t=\infty}log(f(t))^\prime = 0$. Therefore, the term on the second and third line of Eq.~\ref{eq:expanded} becomes 0.

\subsection*{Reconstruction with closed-form estimation of reduced scattering coefficient}
Patterson et al.~\cite{Patterson:89} proposed a closed-form estimate of the reduced scattering coefficient from the time-resolved measurement. The forward model is complex, and the previously proposed method of the optical estimation suffers from model mismatch. Our framework is more robust than closed-form estimation, as our method estimates reduced scattering coefficient by minimizing the L2 loss between the forward model and actual measurement.

\begin{figure}
        \centering
        \includegraphics[width=0.95\linewidth]{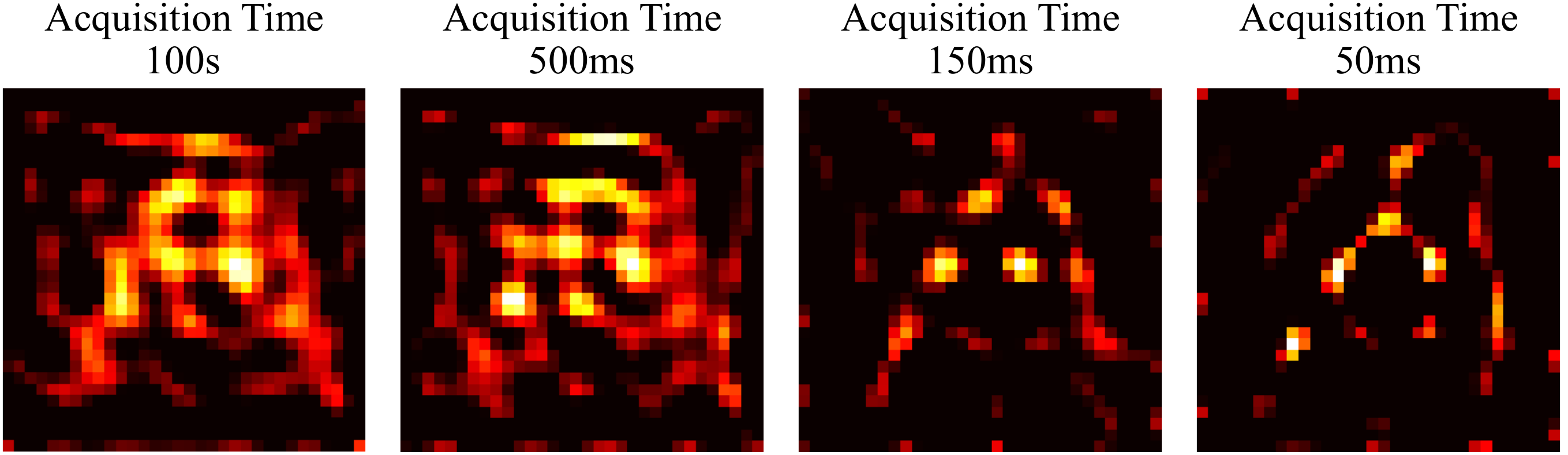}
        \caption{\textbf{Reconstruction results with closed-form estimation of the reduced scattering coefficient.} The estimation of optical properties are not robust to measurement noise, which results in inferior reconstruction quality than the reconstruction with our framework.}
    \label{fig:closed_form_results}
\end{figure}

Fig.~\ref{fig:closed_form_results} illustrates the reconstruction results with closed-form estimation of the optical properties. Because the recovery of the 3D blur kernel is not robust, the quality of the reconstruction is worse than our method (Fig. 6 in the main manuscript).

\subsection*{Reconstruction with Tikhonov Regularizartion} 
Our reconstruction result exploited l1 sparsity and small total variation for the prior. This prior is valid for various practical scenarios, such as fluorescence imaging. Here, we also show reconstruction results with Tikhonov regularization, which is more general:
\begin{equation}
    \hat{\mathbf{v}} = \argmin_\mathbf{v} \frac{1}{2}\norm{\mathbf{b} - \mathbf{Av}}^2 + \lambda \norm{\mathbf{v}}^2.
\end{equation}

With Tikhonov regularization, the target can be estimated as
\begin{equation}
\hat{\mathbf{v}} = (\mathbf{A^\top A} + \lambda I)^{-1} \mathbf{A^\top b}.
\label{eq:Tikhonov}
\end{equation}

\begin{figure}
        \centering
        \includegraphics[width=0.5\linewidth]{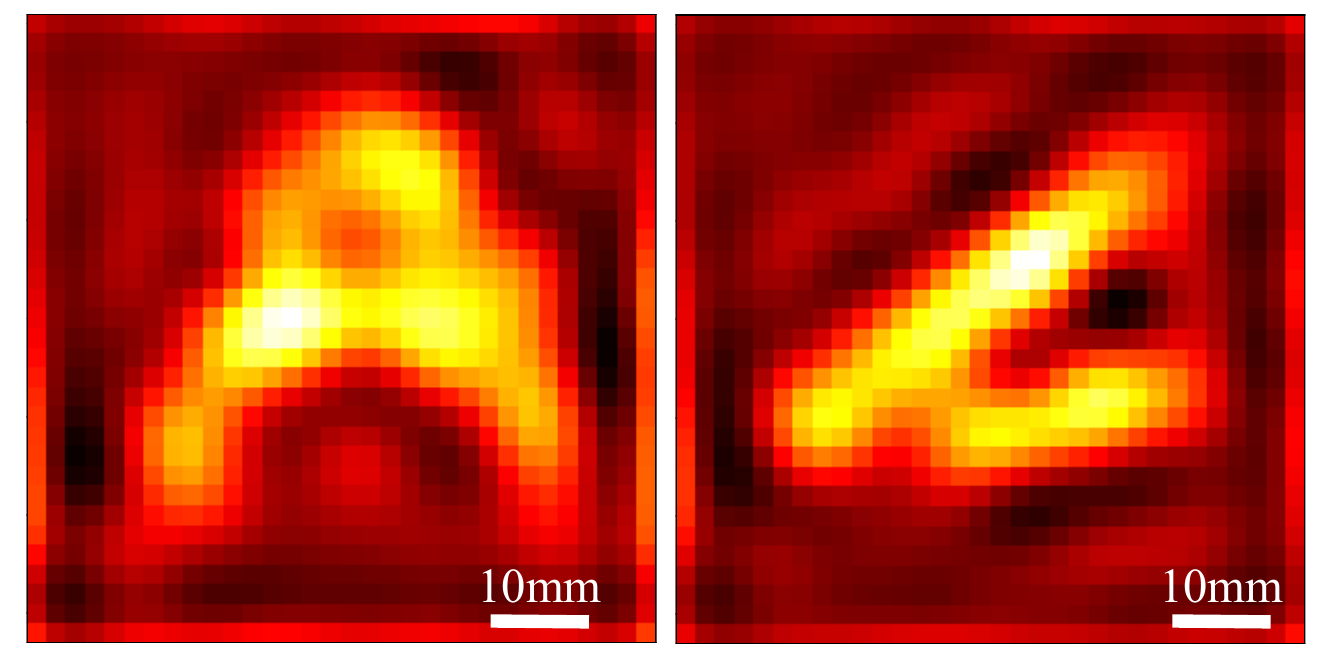}
        \caption{\textbf{Reconstruction results with Tikhonov regularization.} While Tikhonov regularization does not make the reconstruction as sharp as those shown in Fig.6 in the main text, it stills reveals details (e.g., the horizontal line in "A") that are not apparent from the time-averaged image or ballistic frame.}
    \label{fig:tikhonov_results}
\end{figure}

Fig.~\ref{fig:tikhonov_results} shows the reconstruction results of "A" and Wedge targets with $\lambda = 0.1$. Though the reconstruction is more blurry than the results with l1 + total variation regularization, it still reveals details in the target that are not obvious from time-averaged and ballistic frames. For example, the horizontal line of the "A" target is visible in the reconstruction with the Tikhonov regularization.

\end{document}